\documentclass[traditabstract]{aa} 
\usepackage{graphics,color}
\usepackage{epsfig}
\usepackage{txfonts}
\usepackage{natbib}
\usepackage{url}
\usepackage{bm}
\usepackage[latin1]{inputenc} 
\newcommand{\e}[1]{\cdot10^{#1}}
\newcommand{\text}[1]{\mathrm{#1}}
\newcommand{\Fig}{Fig.}

\begin{document}
          \title{The Validity of the Super-Particle Approximation during Planetesimal Formation}
   \titlerunning{The validity of the super-particle approximation}

   \author{Hanno Rein 
          \and
          Geoffroy Lesur 
          \and
          Zoë M. Leinhardt 
          }

   \institute{	University of Cambridge, Department of Applied Mathematics and Theoretical Physics,\\
		Centre for Mathematical Sciences, Wilberforce Road, Cambridge CB3 0WA, UK\\
              \email{hr260@cam.ac.uk}
             }

   \date{Submitted: 11 July 2009 - Revised: 18 December 2009 - Accepted: 22 December 2009 }


\abstract{
The formation mechanism of planetesimals in protoplanetary discs is hotly debated. Currently, the favoured model involves the accumulation of meter-sized objects within a turbulent disc, followed by a phase of gravitational instability. At best, one can simulate a few million particles numerically as opposed to the several trillion meter-sized particles expected in a real protoplanetary disc. Therefore, single particles are often used as super-particles to represent a distribution of many smaller particles. It is assumed that small-scale phenomena do not play a role and particle collisions are not modelled. The super-particle approximation is not always valid when applied to planetesimal formation because the system can be marginally collisional (of order one collision per particle per orbit). The super-particle approximation can only be valid in a collisionless or strongly collisional system, although, in many recent numerical simulations this is not the case.

In this work, we present new results from numerical simulations of planetesimal formation via gravitational instability. 
A scaled system is studied that does not require the use of super-particles. This system is simplified for computational practicality and proper identification of important processes:
1) the evolution of particles is studied in a local shearing box; 2) the particle-particle interactions such as gravity, physical collisions, and gas drag are solved directly assuming a constant background shear flow without any feedback from the particles. We find that the scaled particles can be used to model the initial phases of clumping if the properties of the scaled particles are chosen such that all important timescales in the system are equivalent to what is expected in a real protoplanetary disc. Constraints are given for  the number of particles needed in order to achieve numerical convergence.  

We compare this new method to the standard super-particle approach. We find that the super-particle approach produces unreliable results that depend on artifacts such as the gravitational softening in both the requirement for gravitational collapse and the resulting clump statistics.  
Our results show that short-range interactions (collisions) have to be modelled properly.
}

\maketitle

\section{Overview}\label{sec:overview}

Extrasolar planets have been observed around a variety of parent stars from pulsars to solar-type stars to M-dwarfs \cite[see e.g.,][]{directimaging,Wol1992} indicating that planet formation is common and successful in a broad range of environments. However, the process of planet formation itself is not directly observable, leaving theory and numerical simulations to fill in the blanks between observations of hot circumstellar discs around young stars and planets orbiting mature stars. One of the most important unanswered questions in the theory of planet formation is what is the mechanism for planetesimal formation, i.e., the process by which the building blocks of planets are formed. 

There are two main theories for planetesimal formation: mutual collisions \citep[e.g.,][]{Hayashi77} and gravitational instability in the dust layer \citep{GoldreichWard1973}. In the first hypothesis, dust particles grow as the result of accretion-dominated collisions. Although the formation of planetesimals by mutual collisions is consistent with meteoritic evidence, the collision speed between dust particles (or aggregates) must be much slower than the typical velocity dispersion in a standard protostellar disc to avoid destructive collisions \citep{BlumWurm2008}. In addition, the planetesimal formation process is so slow that meter-sized particles are in danger of spiraling into the star before growing large enough to decouple from the gas \citep{Weidenschilling1977}. Even if km-sized planetesimals were able to form, they would be in danger of being ground down again by mutual collisions \citep{Stewart2009}. 

Gavitational instability is often considered to be a solution to most of these problems because the intermediate sizes are avoided all together. In this theory, the dust layer becomes dense enough for the Keplerian shear and velocity dispersion of the dust particles to be unable to support the dust against its own self gravity. The dust then collapses into clumps that eventually cool via drag forces and mutual collisions into planetesimals. 
Many different authors have worked on this subject. Until recently, the focus has been on quiet, non-turbulent, and low density regions of the accretion disc \citep[see e.g.,][]{Michikoshi2009,Michikoshi2007,Tanga2004}.
However, the turbulent gas in the protoplanetary nebula stirs the dust, which increases the velocity dispersion of the dust particles. Several ideas have been proposed to overcome the turbulence-induced mixing of the dust particles and create localized clumps. For example, \citet{Cuzzi2008} suggest that the same turbulence that stirs the dust on larger scales may also collect the dust particles on small scales. A similar idea was proposed by \citet{Johansen2007}, in which dust particles are localized into clumps using both turbulence and the streaming instability \citep{YoudinGoodman2005}. The clumps then become gravitationally unstable.  A third hypothesis suggests that large structures, such as vortices, may be able to collect and protect dust particles from the turbulent background \citep[][]{BargeSommeria95}.
In this paper, we focus on the gravitational collapse in a very dense and turbulent region of the protoplanetary disc. The overdensity in the dust layer, which is approximately two orders of magnitude higher than the standard minimum-mass solar nebula, might have been created by any of the processes described above.

Numerical simulations must be used to test models of planetesimal formation. However, the separation of scales in the problem is huge. A meter-sized object is more than 12~orders of magnitude smaller than the protoplanetary disc thickness forcing numericists to use \textit{super-particles}. These super-particles have a mass much higher than the mass of a single dust particle, but the forces (e.g., drag forces) acting on them are equivalent to those of a single dust particle. \cite{Tanga2004} were able to use this approximation to succesfully model the gravitational collapse in a regime in which collisions are unimportant. However, when the surface density is increased by two orders of magnitude, as mentioned above, the system does become marginally collisional and therefore the super-particle approach breaks down. 

\cite{Michikoshi2009} perform a similar set of simulations, also in a low density, non-turbulent region of the disc. Although the authors describe their particles as super-particles, the way in which they accurately treat the collisions is the same as the method presented in this paper. Other approaches use Monte Carlo type methods to resolve statistically the outcome of overlapping particles \citep[see e.g.,][]{Lithwick2007}.  

In the following, we look carefully at the numerical requirements of modelling gravitational instability accurately and test the validity of using super-particles in a high density region. For simplicity, we begin with a system without turbulence and assume that the surface density has already been increased by turbulence or some equivalent process. We find that the super-particle approach gives erratic results when collisions become important.

We therefore go on and present a method that does not use super-particles. We simulate a scaled system in which the number of particles is dramatically lower than in a real protoplanetary disc. To keep the surface density the same, the mass of individual particles has to be increased. The drag force produced by the surrounding gas is scaled accordingly, so that the stopping time remains constant. Furthermore, we increase the physical size of each particle. This allows us to keep the collision timescales in simulations of different particle numbers exactly the same. These requirements lead to results that are independent of the number of particles, which is the only free parameter in the simulation, and we call those simulations converged. The results agree with other simulations performed in the context of Saturn's rings \citep{GoldreichTremaine1982,DaisakaIda1999}. In this paper, we argue that this method should also be used in current simulations of planetesimal formation.

The outline of this paper is as follows.
In Sect.~\ref{sec:def}, we define the important timescales in the problem and show that the super-particle approach breaks down in high density regions. In Sect.~\ref{sec:methods}, we discuss the numerical method and the initial conditions of our simulations. Results of numerical simulations using both the super-particle approach and our scaling method are presented in Sect.~\ref{sec:results}. We conclude the paper with resolution constraints for numerical simulations and discuss the implications for the planetesimal formation process through gravitational instability in Sect.~\ref{sec:disc}.

\section{Orders of magnitude}\label{sec:def}
\subsection{Definition of timescales}
The aim of this work is to simulate the dynamics of dust (or particles) interacting gravitationally inside a protoplanetary disc. 
Each particle is subject to five different physical processes, each operating on various typical timescales:

\begin{description}
\item \textit{Stopping time} $\tau_s$
  
 Each particle feels the effects of the surrounding gas through a linear drag force. This force can be written as $\mathbf{F}_\text{drag}=\frac{m_p}{\tau_s} (\mathbf{v}_g-\mathbf{v}_p)$, where $\tau_s$ is the stopping time of a particle of mass $m_p$, $\mathbf{v}_g$ is the velocity of the gas, and $\mathbf{v}_p$ is the velocity of the particle \citep{W77}.\\
 
\item \textit{Physical collision timescale} $\tau_c$ 

Dust particles will suffer a physical collision with another dust particle on a timescale of $\tau_c=(\sigma_c \mathrm{\bar v}_p n)^{-1}$, where $\sigma_c$ is the geometrical cross-section of the particles, $\mathrm{\bar v}_p$ is the particle velocity dispersion, and $n$ is the number density of particles. \\

\item \textit{Orbital timescale} $\tau_e$ 

The timescale associated with the particles orbiting the central object in a local shearing patch is the epicyclic period $\tau_{e}=2\pi\Omega^{-1}$, where $\Omega$ is the angular velocity at the semi-major axis of the shearing box.

\end{description}

\noindent In this paper, we consider two limits for the gravitational interactions between dust particles, long-range and short-range interactions. The long-range interaction can be seen as a collective process involving interactions between clouds of particles. On the other hand, the short-range interaction is important for resolving close approaches between pairs of particles (i.e., gravitational scattering). For the sake of clarity, we separate these two processes. 

\begin{description}
\item \textit{Gravitational collapse timescale} $\tau_{Gl}$ 

We introduce a length scale $\lambda \gg \delta_r$, where $\delta_r$ is the average distance between neighbouring dust particles. In this limit, which we call the long-range interaction, the dust particle distribution can be approximated by a continuous density $\rho$, and one can define a gravitational timescale, which is of the order of the free fall time of the system, defined by $\tau_{Gl}=1/\sqrt{G \rho}$ where $G$ is the gravitational constant. \\

\item \textit{Gravitational scattering timescale} $\tau_{Gs}$ 

Short-range gravitational interactions can be seen as an interaction between a pair of single particles that can result in a scattering event. As with physical collisions, the timescale for a gravitational interaction depends on the cross-section, $\tau_{Gs}=(\sigma_G \mathrm{\bar v}_p n)^{-1}$, where $\sigma_G$ is the gravitational cross-section of the particle. Using Kepler's laws, one finds that $\sigma_G\sim G^2 m_p^2/\mathrm{\bar v}_p^4$. The gravitational cross-section is velocity dependent, which makes it qualitatively different from physical (billard ball) collisions between two particles. We note that gravitational focusing can also lower the physical collision timescale.

\end{description}

\noindent As mentioned earlier (Sect. \ref{sec:overview}), the particles might also be affected by excitation from a turbulent background state on a timescale defined by the turbelence itself. 
In this paper, we model this excitation in a simplified way, described in Sect. \ref{sec:methods}. In our case, the turbulence is scale independent and has no timescale associated with it. 

\subsection{The real physical system}\label{sec:real}

To identify the dominant physical processes in a protoplanetary disc, one has to quantify and compare the relevant timescales as defined above. 
In the following discussion and in the numerical simulations, we assume $R_0=1$ AU, a gas disc thickness of $H/R_0=0.01$, and a minimum mass solar nebula (MMSN) with surface gas density $\Sigma=890~\text{g~cm}^{-2}$. 
One usually assumes a solid to gas ratio of~$\rho_s/\rho_g=0.01$. We, however, assume a solid to gas ratio of unity. As mentioned above, this value is justified by numerical simulations \citep{Johansen2007,Johansen2009} that demonstrate overdensities of the order of~$10^1-10^2$ can easily occur because of the interaction with a turbulent gas disc, the streaming instability, or vertical settling.
We note that significantly different models of the solar nebula have been proposed \citep{Desch2007}. However, all our results can easily be scaled to different scenarios.

We simplify the system by assuming that all solid components of the disc are in meter-sized particles. We assume that these boulders have a typical velocity dispersion caused by gas turbulence $\mathrm{\bar v}_p\sim0.05 c_s\sim 30\text{m}/\text{s}$, where $c_s$ is the local sound speed, in accordance with numerical results \citep{Johansen2007}. Since $\mathrm{\bar v}_p\ll c_s$, the particles tend to sediment toward the midplane, forming a finite thickness dust layer due to the non-zero velocity dispersion.

For meter-sized boulders, the physical cross-section is $\sigma_c\sim 1\,\mathrm{m}^2$, whereas the gravitational scattering cross-section is $\sigma_G\sim10^{-21}\,\mathrm{m}^2$ showing clearly that gravitational scattering is irrelevant to the dynamics of dust particles embedded in a disc. 
However, all other timescales are roughly equivalent. 
Meter-sized boulders are weakly coupled to the gas with a stopping time $\tau_s\sim \tau_e\sim \Omega^{-1}$ \citep{W77}. 
The physical collision timescale is $\tau_c=7.3 \, \Omega^{-1}$ and the long-range gravitational interaction timescales is $\tau_{Gl} \sim \Omega^{-1}$.

This physical system is expected to become gravitationally unstable, according to the Toomre criterion \citep{Toomre1964}.
The instability occurs when the gravitational collapse timescale $\tau_{Gl}$ is shorter than the transit time due to random particle motions $\lambda /\mathrm{\bar v}_p $ and the orbital timescale. 
Assuming that the particles can be modelled by an isotropic gas\footnote{This would be true for a strongly collisional system, but is not formally valid in the studied regime.} with a sound speed $\mathrm{\bar v}_p$, the system becomes unstable when
\begin{eqnarray}
Q\equiv \frac{\mathrm{\bar v}_p \Omega}{\pi G \Sigma}<Q_\mathrm{crit}\simeq 1.\label{eq:toomre}
\end{eqnarray}
In that case, the most unstable wavelength is given by
\begin{eqnarray}
\lambda_T&=&\frac{2\pi^2G\Sigma}{\Omega^2}. \label{eq:toomrelength}
\end{eqnarray}
In the following, we compare the physical parameters from this section to their counterpart in numerical simulations. We first summarize the super-particle approach before presenting the scaling method.

\subsection{Super-particle approximation}
A super-particle represents many smaller particles. The dynamics of the small particles are not calculated exactly. It is assumed that all particles behave similarly and in a collective manner. To simulate gravitational interaction between super-particles, one has to use a softened potential. Without this, the gravitational scattering cross-section of super-particles becomes too large and super-particles undergo gravitational scattering events, which is unphysical since these events never occur in a real system. Individual particle-particle collisions are not modelled.

There are various examples where this approach is used successfully. For example, smoothed particle hydrodynamics (SPH) uses the super-particle approximation to simulate many gas molecules \citep{Lucy1977}. These systems are often assumed to be \emph{strongly collisional} to ensure thermodynamical equilibrium inside each super-particle. One can therefore assign collective properties to clouds of particles such as pressure and temperature. Another example is the evolution of galaxies. When two galaxies collide, individual particles (stars) will usually not undergo gravitational scattering events or physical collisions. In that case, the super-particle approach models a \emph{collisionless} system in which collective dynamics is the only important physical process. In both cases, the super-particle approximation is a valid approach for simulating the system numerically, but will break down as soon as the system is \emph{marginally} collisional.

As an example, we consider two clouds of particles undergoing a ``collision''. In the strongly collisional case, the clouds will slowly merge and the thermodynamic variables (e.g., temperature) will diffuse between the clouds, the particles inside each cloud following a random walk trajectory due to numerous collisions. On the other hand, in the collionless regime, the two clouds simply do not see each other because there is no short-range interaction present between the particles. In a marginally collisional regime, some particles will collide with particles of the other cloud, leading to a partial thermalisation of the velocity distribution, but some other particles will not have any collision at all and will follow approximately a straight line. Evidently, the outcome of this event \emph{cannot} be described using a super-particle approach.

\subsection{Scaling method}
The idea of our scaling method is that one should keep all important timescales in a numerical simulation as close to those of the real physical system as possible and model all particle collisions explicitly. 

The numerical system consists of $N_\mathrm{num}$ particles in a box with length $H$, simulating a patch of the disc at a fixed radius $R_0$. 
The particle mass is $m_\mathrm{num}=\Sigma H^2/N_\mathrm{num}$ ($H$ is the box size and one scale height). Thus, the density in the box and the long-range gravitational interactions are unchanged compared to the real system.

The gravitational scattering cross-section of the numerical system is then given by 
\begin{eqnarray}
\sigma_{G,\mathrm{num}}=\frac{G^2 \Sigma^2 H^4}{N_\mathrm{num}^2 \mathrm{\bar v}_p^4} \label{eq:crossgravity}.
\end{eqnarray}
For the initial surface density and velocity dispersion used in the simulation (Sect. \ref{sec:real}), one finds
\begin{eqnarray}
 \sigma_{G,\mathrm{num}}\simeq \frac{ H^2}{N_\mathrm{num}^2} & \simeq & \frac{0.0001\text{ AU}^2}{N_{\mathrm{num}}^2}.
\end{eqnarray}
The physical collision cross-section in the simulation is $\sigma_{c,\mathrm{num}}~=~\pi a_\mathrm{num}^2$ where $a_\mathrm{num}$, is the radius of the particles in the simulation. 
We derive two constraints from the physical and gravitational collision cross-sections:
\begin{enumerate}
\item \textit{Same mean free path in simulation and real physical systems}\nopagebreak

That means $N_{\mathrm{num}} \sigma_{c,\mathrm{num}} =N  \sigma_c $, where $\sigma_c$ and $N$ are the geometrical cross-section and the number of particles in a region of size $H^3$ in a real disc, respectively.
In other words, this condition ensures that the physical collision timescale in the simulation is exactly the same as in the real system. 
\\

\item \textit{Negligible gravitational scattering}\nopagebreak

 When two particles approach each other, the outcome should be a physical collision, which means that $\sigma_{c,\mathrm{num}}~\gg~\sigma_{G,\mathrm{num}}$.
The gravitational scattering timescale remains long compared to the physical collision timescale.
\end{enumerate}

\noindent The first condition places a constraint on the particle size in the simulation. With $N=4.7\cdot 10^{18}$ and a particle radius of $a=1\,\mathrm{m}$ as found in a real disc within a box of volume $H^3$, using the parameters from above finds that
\begin{eqnarray}
  a_\mathrm{num}  = \sqrt{\frac{\sigma_{c,\mathrm{num}}}\pi} =  \sqrt{\frac{N\sigma_c}{N_{\mathrm{num}}\pi}} \simeq \frac{0.014}{\sqrt{N_\mathrm{num}}}\mathrm{AU}.\label{eq:anum}
\end{eqnarray}
We note that once this condition is satisfied in the initial conditions it will be automatically satisfied at all times. In simulations, we typically find that the collision timescale is reduced by more than one order of magnitude during the gravitational collapse.

The second condition is then satisfied by changing the number of particles.
An interesting result is that if $N_\mathrm{num}>10$, the second condition is easily satisfied if the first one is satisfied. 
We note however that $\sigma_{G,\mathrm{num}}$ depends strongly on the velocity dispersion. 
In particular, a velocity dispersion 5 times smaller than the initial value (as found in some simulations) leads to an increase by a factor of 600 in the gravitational scattering cross-section. 
Therefore, we suggest using a large safety factor ($N_\mathrm{num}>10^5$) to ensure that the second condition is always satisfied, even for significantly smaller~$\mathrm{\bar v}_p$. This condition also allows us to estimate when our approach breaks down, namely when the number of clumps in the system is so low ($N<10\sim100$) that gravitational scattering becomes important.

Although it would be helpful as a further simplification, it is not possible to perfom the above mentioned simulation in two dimensions. In that case, the filling factor, which is defined to be the ratio of the volume (or area) of all particles to the total volume (or area), is of the order of one for the above parameters. We note that in 2D an increase in particle number (and decrease in particle size as required by Eq. \ref{eq:anum}) does not decrease the filling factor if the collisional lifetime remains constant.

\section{Methods}
\label{sec:methods}

Two different kinds of simulation are considered in this paper: 
\begin{enumerate}
  \item Particles are assumed to be point masses and have no physical size, the gravitational field of the particles being approximated with a smoothing length to avoid numerical divergences (see Sect.~\ref{sec:nbody}).

  \item Particles have a physical size and, therefore, no gravitational smoothing is required but physical collisions must be included. 
\end{enumerate}
We refer to the particles as super-particles and scaled particles, respectively.

\begin{figure}[tbh]
\center
\resizebox{\columnwidth}{!}{ 
\begin{picture}(0,0)%
\epsfig{file=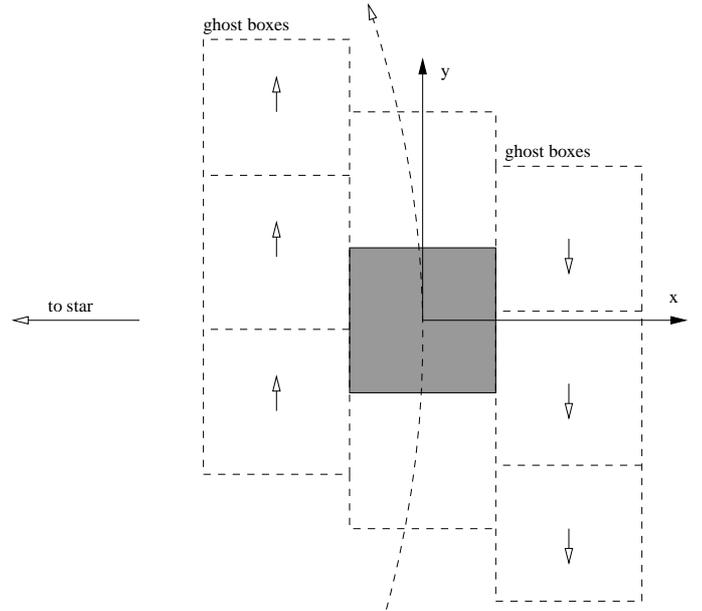}%
\end{picture}%
\setlength{\unitlength}{3947sp}%
\begingroup\makeatletter\ifx\SetFigFont\undefined%
\gdef\SetFigFont#1#2#3#4#5{%
\fontfamily{#3}\fontseries{#4}\fontshape{#5}%
\selectfont}%
\fi\endgroup%
\begin{picture}(5574,5041)(739,-5844)
\put(4801,-2086){\makebox(0,0)[lb]{\smash{{\SetFigFont{12}{14.4}{\familydefault}{\mddefault}{\updefault}{\color[rgb]{0,0,0}ghost boxes}%
}}}}
\put(4276,-1411){\makebox(0,0)[lb]{\smash{{\SetFigFont{12}{14.4}{\familydefault}{\mddefault}{\updefault}{\color[rgb]{0,0,0}y}%
}}}}
\put(6151,-3286){\makebox(0,0)[lb]{\smash{{\SetFigFont{12}{14.4}{\familydefault}{\mddefault}{\updefault}{\color[rgb]{0,0,0}x}%
}}}}
\put(1051,-3361){\makebox(0,0)[lb]{\smash{{\SetFigFont{12}{14.4}{\familydefault}{\mddefault}{\updefault}{\color[rgb]{0,0,0}to star}%
}}}}
\put(2326,-1036){\makebox(0,0)[lb]{\smash{{\SetFigFont{12}{14.4}{\familydefault}{\mddefault}{\updefault}{\color[rgb]{0,0,0}ghost boxes}%
}}}}
\end{picture}%
}
	\caption{Shearing box, simulating a small patch of the protoplanetary disc. The grey box represents the shearing box of interest, the dashed boxes surrounding the central box represent one ghost ring.} 
	\label{fig:shearingpatch}
\end{figure}

We perform our simulations in a cubic box with shear periodic boundary conditions in the radial ($x$) and perdiodic boundary conditions in the azimuthal ($y$) and vertical ($z$) directions, as illustrated in \Fig \ref{fig:shearingpatch} \citep{Wisdom1988}. 
In the local approximation, the force per mass on each particle is a sum of the contributions from Hill's equations and the interaction terms.
Hill's equation can be written as
\begin{eqnarray}
\bm{F}_\text{hill} &=&-2\Omega \;\mathbf{e}_z\times \mathbf{v}_p+3 \Omega^2 \;x\;\mathbf{e}_x-\Omega^2z\;\mathbf{e}_z. 
\end{eqnarray}
The interaction term $\bm{F}_\text{int}$ is divided into components related to self gravity, physical collisions between particles, or drag and excitation forces:
\begin{eqnarray}
\bm{F}_\text{int} &=& \bm{F}_\text{grav} + \bm{F}_\text{col} + \bm{F}_\text{drag} + \bm{F}_\text{turb}.
\end{eqnarray}
We solve the resulting equations of motion with a leap frog (kick drift kick) time stepping scheme. 
In the following subsections, we describe the numerical methods used to compute each of these terms and their physical relevance.

The box size of $0.01~\text{AU}$ was chosen such that there are always several unstable modes in the box, when the system is pushed into the regime of gravitational collapse, as estimated by Eq.~\ref{eq:toomrelength}. 

\subsection{Self gravity}\label{sec:nbody}
In the $N$-body problem that we consider, we have to solve Newton's equations of universal gravitation for a large number of particles $N$. The gravitational force on the $i$-th particle is given by 
\begin{eqnarray}
\bm{F}_\text{grav} = \sum_{j\neq i} G \frac{m_im_j}{\left( r_{ij} +b \right)^2}\mathbf{e}_{ij}, \label{eq:nbody}
\end{eqnarray}
where $b$ is the smoothing length used to avoid divergences in numerical simulations (in simulations that include physical collisions $b = 0$) and $\mathbf{e}_{ij}$ is the unit vector in the direction of the gravitational force between the $i-$th  and $j-$th particle. 
Calculating the gravity for each particle from each other particle results in $O\left(N^2\right)$ operations. To reduce the number of operations, one can use different approximations to Eq.~\ref{eq:nbody}. In this paper, we use a Barnes-Hut (BH) tree code \citep{Barnes1986}.

The BH tree in three dimensions divides the original box into eight smaller cells with half the length of the original box. 
This process is continued recursively until there is only one particle per cell left. 
The depth of the tree in a homogeneous medium is approximately $\log_8 N$. 
One then calculates the total mass and the centre of mass for every cell at every level of the tree. 

To calculate the force acting on a particle, one starts at the top of the tree and descends into the tree as far as necessary to achieve a given accuracy. 
If the current cell is far away from the particle for which the force is calculated, then the detailed density structure within this cell is not important. 
All that matters is the box's monopole moment (total mass and centre of mass). 
One therefore does not have to descend into this tree branch any further. 
The BH tree reduces the number of calculations to $O\left(N\log N\right)$. 

We use~$8$ \textit{rings} of ghost boxes in the radial and azimuthal direction (\Fig \ref{fig:shearingpatch} shows one ring). A ghost box is simply a shifted copy of the main box. 
The gravity on each particle is then calculated by summing over contributions from each (ghost) box. 
This setup approximates a medium of infinite horizontal extent and avoids large force discontinuities at the boundaries.
We do not need any ghost boxes in the vertical direction because the disc is stratified.

A finite number of ghost rings can only act as an approximation of an infinite medium and the gravitational force will tend to concentrate particles horizontally in the centre of the box. 
Due to the 8 ghost rings, the asymmetry of the gravitational force between the centre and the faces of the box is reduced by a factor of 20 compared to using no ghost rings at all. We note, however, that a small
asymmetry remains in our simulations and leads to higher concentrations in the box centre.
In other simulations such as \cite{Tanga2004}, this effect is not seen because the system is initially gravitationally unstable and integrated only for approximately one orbit. 
The system that we are interested in is marginal gravitationally unstable and this slight asymmetry will become important after several orbits. In the beginning of our simulations, we integrate for many orbits in order to reach a stable equilibrium. We then push the system into a gravitationally unstable regime (see Sect.~\ref{sec:initcond} for details). During the stable phase, horizontal over-concentrations are to be expected and are indeed observed because of this asymmetry.
However, since in a real protoplanetary disc the gravitational instability is considered to appear locally (e.g., inside a vortex or a similar structure), this effect of having a preferred location for gravitational instability is not unphysical and does not affect any conclusions made in this paper. We tested this by applying a linear cut-off to the gravitational force at a distance of one box length (A. Toomre, private communication). In this case, there was no preferred location in the box and the simulation evolved in exactly the same way.

\subsection{Physical collisions}
Physical collisions between particles are treated in the following way. After each timestep, we check whether any two particles are overlapping. Using the already existing tree structure from the gravity calculation, one can again reduce the computational costs from $O\left(N^2\right)$ to $O\left(N \log N\right)$. In these simulations, a collision between particles is defined as an overlap between two particles that are approaching each other. When a collision is detected, it is resolved assuming energy and momentum conservation (perfectly elastic collisions). We note that the timestep has to be small enough such that no collision is missed and the overlap is always small compared to the particle size. 

The particle radius is given by Eq. \ref{eq:anum}. In order to keep the collision timescale $\tau_c$ close to unity, which is numerically the worst case scenario because all timescales are equivalent, we increase the radius by a factor of 4 in all simulations.

\subsection{Drag force}
Each particle feels a drag force. The background velocity of the gas $\mathbf{v}_g$ is assumed to be a steady Keplerian profile
\begin{eqnarray}
\mathbf{v}_g=-\frac32 \Omega x\;\mathbf{e}_y.
\label{gasvel}
\end{eqnarray}
Although accretion flows are turbulent, we choose to use a simple velocity profile so that the behaviour of self-gravitating particles can be understood in conditions that can be easily controlled.

\subsection{Random excitation}

The system of particles described above is dynamically unstable even without self-gravity, as the coefficient of restitution and the stopping time do not depend on the particle velocities\footnote{This instability is qualitatively similar to the instability described by \cite{GoldreichTremaine1978a} for Saturn rings, although in the latter case the drag forces are absent.}. 
The particles can either settle down in the midplane and create a razor-thin disc if the stopping time is too short, or they can expand vertically forever if the stopping time is above some critical value and the excitation mechanism is provided only by collisions. Dust particles in accretion discs are found in the settling regime. However, a complete settling never occurs in accretion disks because the background flow is always turbulent due to the Kelvin-Helmoltz instability \citep{JohansenHenning2006}, the MRI, or other hydrodynamic instabilities \citep[see e.g.,][]{LesurPapaloizou2009} which diffuse particles vertically \citep{FromangPapaloizou2006}. This stirring process of turbulence is not present in the gas velocity field of the simulations presented here (see Eq.~\ref{gasvel}). To approximate the turbulent mixing, we added a random excitation (white noise in space and time) to the particles in the simulation. This allows us to have a well defined equilibrium in which the system is stable rather than starting from unstable initial conditions that might influence the final state.

We perturb the velocity components of each particle after each timestep on a scale $\Delta \mathrm{v}_i =\sqrt{\delta t}\, \xi$, where $\delta t$ is the current timestep and $\xi$ is a random variable with a normal distribution around $0$ and variance $s$. This excitation mechanism \emph{heats up} the particles and allows us to have a well defined three dimensional equilibrium as shown in Sect.~\ref{sec:results}. In our simulations, we use a value of $s=1.3\e{15}\;\text{m}^2\text{s}^{-3}$ and $s=8.9\e{14}\;\text{m}^2\text{s}^{-3}$ for the simulations without and with collisions, respectively. These values were chosen such that the equilibrium state is approximately equivalent for both types of simulations.

\subsection{Initial conditions}\label{sec:initcond}

All particles, which have equal mass, smoothing length, and physical size, are placed randomly inside the box in the $x-y$~plane. We note that particles have either a physical size or a smoothing length associated with them, depending on the type of simulation we perform (Sect.~\ref{sec:methods}). In the z-direction, the particles are placed in a layer with an initially Gaussian distribution about the midplane and a standard deviation of 0.05 H. We allow the system to reach equilibrium by integrating it until $t=30\Omega^{-1}$ ($4.8\text{yrs}$). 

We tested various ways of pushing the system into the unstable regime, either by reducing the turbulence stirring or by shortening the stopping time, but did not see any qualitative difference as long as we start from a well defined equilibrium state. In the simulations presented in this paper, we switch off the turbulence stirring. Following this modification, the system becomes gravitationally unstable and bound clumps form within a few orbits.

\section{Results}
\label{sec:results}

\subsection{Super particles}
\begin{figure}[btp]
\centering
\includegraphics[angle=270,width=\columnwidth]{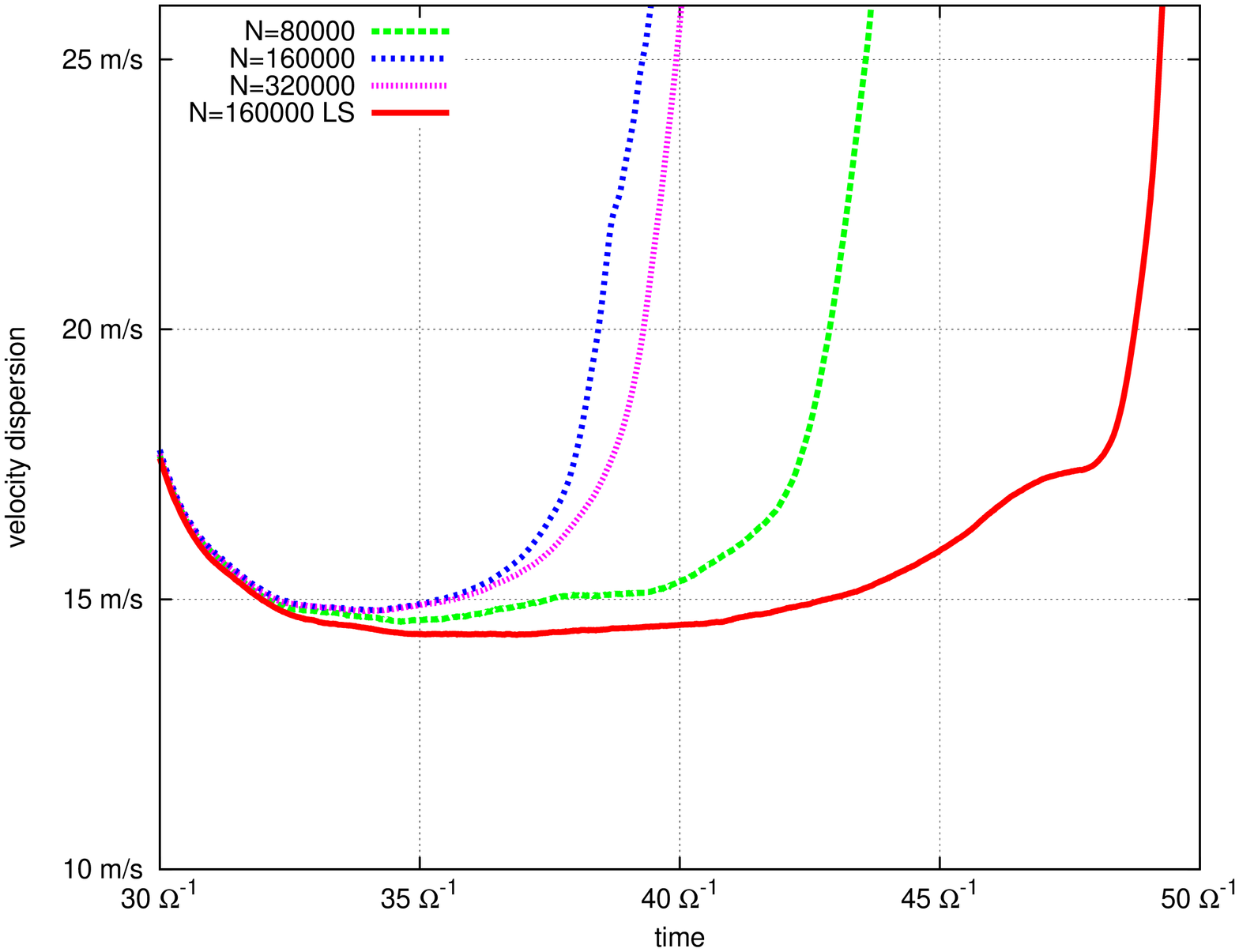}
\caption{Super-particles: Velocity dispersion as a function of time in four different runs. The simulation labeled \texttt{LS} has a ten times larger smoothing length. The curves do not overlap because the simulations are not converged.  \label{fig:veldisp_smooth}} 
\includegraphics[angle=270,width=\columnwidth]{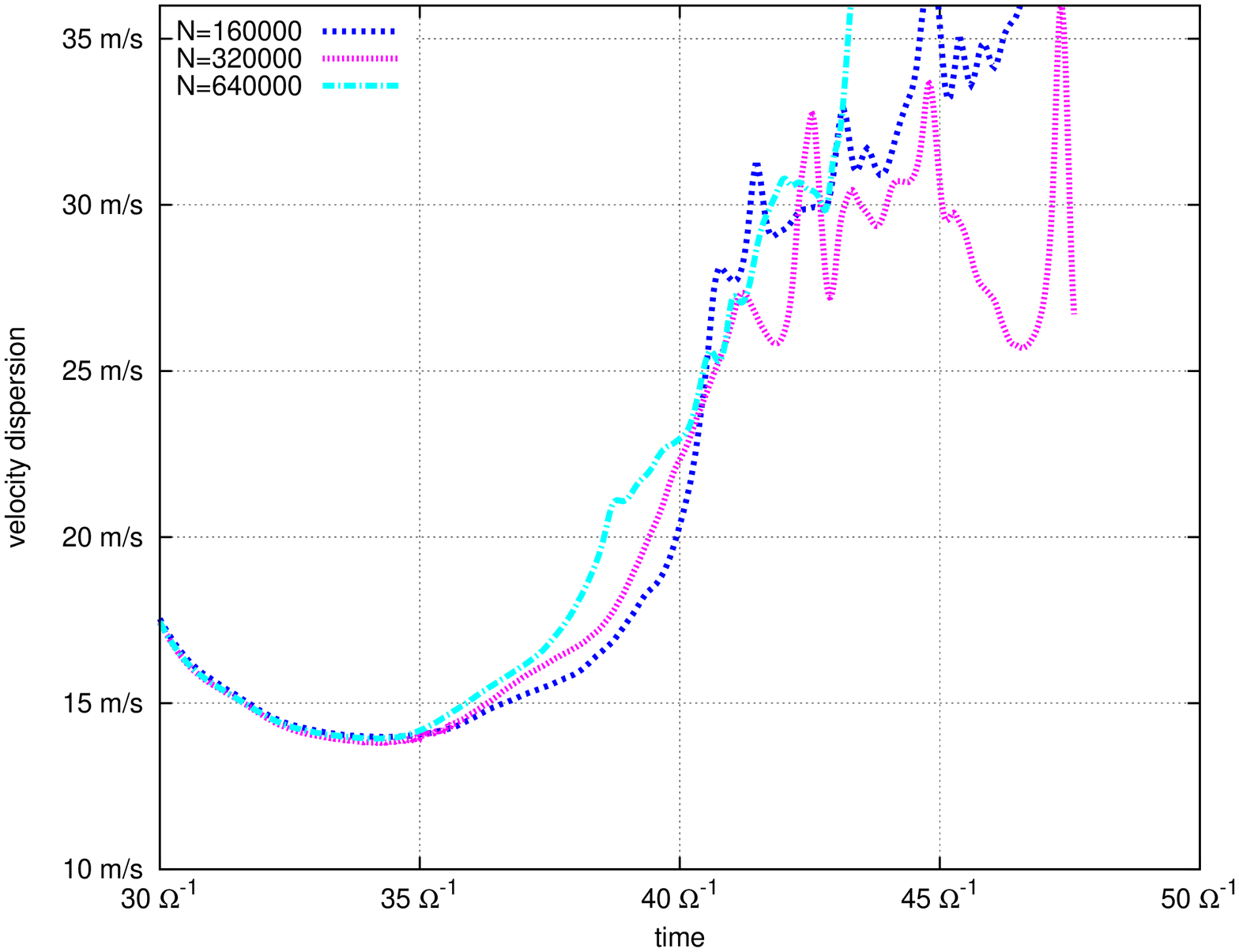}
\caption{Scaled-particles: Velocity dispersion as a function of time in three different runs that include physical collisions. All curves overlap because the simulations are converged. \label{fig:veldisp_scaled} }
\end{figure}

We first present simulations without physical collisions (super-particles) which rely on the smoothing length $b$ to avoid any divergencies. Although we use a tree code and therefore a smoothed potential for the force calculation is assumed, the results are equivalent to an FFT-based method where the grid length acts as an effective smoothing length. In general, to check the numerical resolution, the particle number $N_\text{num}$ is increased and the smoothing length $b$ is reduced independently. A simulation is resolved when the result is independent of both $N_\text{num}$ and $b$. This turns out to be impossible in the present situation. The main reason is that the smallest scale in a gravitational collapse will ultimately depend on the smoothing length. By varying both parameters at the same time, an empirical scaling of~$b\sim1/\sqrt{N_\text{num}}$ works fairly well if $N_\text{num}$ is large enough. However, this procedure is not justified and is unphysical. The smoothing length was introduced to avoid divergent terms and not to model any small-scale physical process. Therefore, it should \emph{not} have any impact on the physical outcome of the simulation. Incidentally, the existence of this dependency indicates that \emph{short-range interactions are important} for the result of these simulations and should be modelled with care.

In \Fig \ref{fig:veldisp_smooth}, we plot the velocity dispersion as a function of time. The simulations begin from the stable equilibrium described in Sect.~\ref{sec:methods}. After $t=30~\Omega^{-1}$, we switch off the excitation mechanism. Because the system continues to cool, it becomes gravitationally unstable within one orbit, as estimated by {Eq.~\ref{eq:toomre}}.
Once clumping occurs, the velocity dispersion begins to rise again. All simulations use the particle radius given by {Eq.~\ref{eq:anum}} as a smoothing length except the ones labeled \texttt{LS}, which use a ten times larger value. The larger softening length is approximately a $128$-th of the box length and illustrates the kind of evolution expected from an FFT method using a $128^3$ grid.  

\begin{figure*}[p]
\centering
\includegraphics[width=0.9\textwidth]{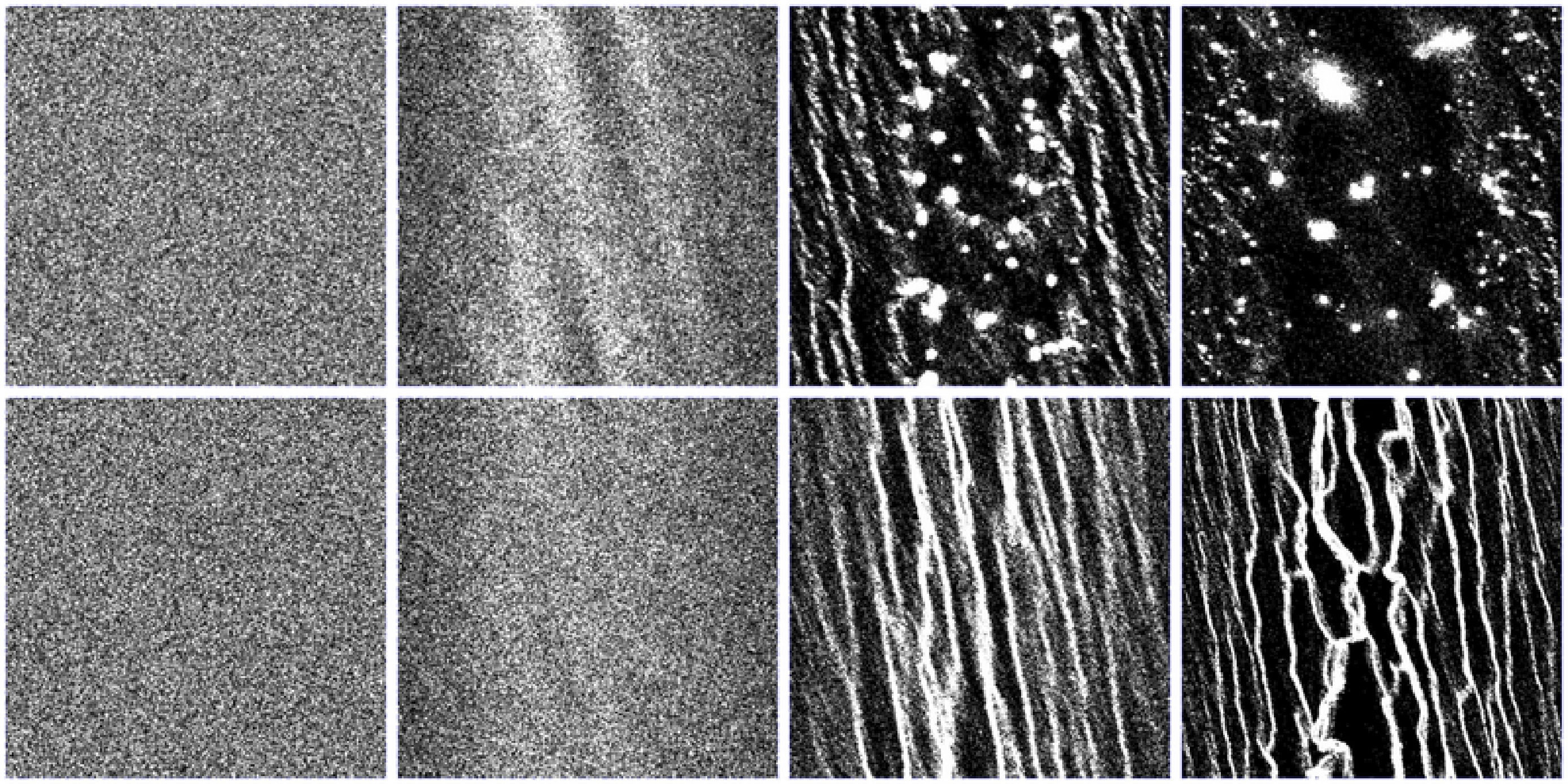}
\caption{Super-particles: Snapshots of the particle distribution in the $xy$ plane. Both simulations use 160~000 particles with different smoothing lengths. The simulation on the bottom uses a ten times larger smoothing length than the one on the top. The snapshots were taken (from left to right) at $t=0, 30, 37, 40\,\Omega^{-1}$. With a large smoothing length, the outcome looks very different, the system is more stable, more stripy structure can be seen and clumps form later, if at all. \label{fig:snapsmoothing}}
\includegraphics[width=0.9\textwidth]{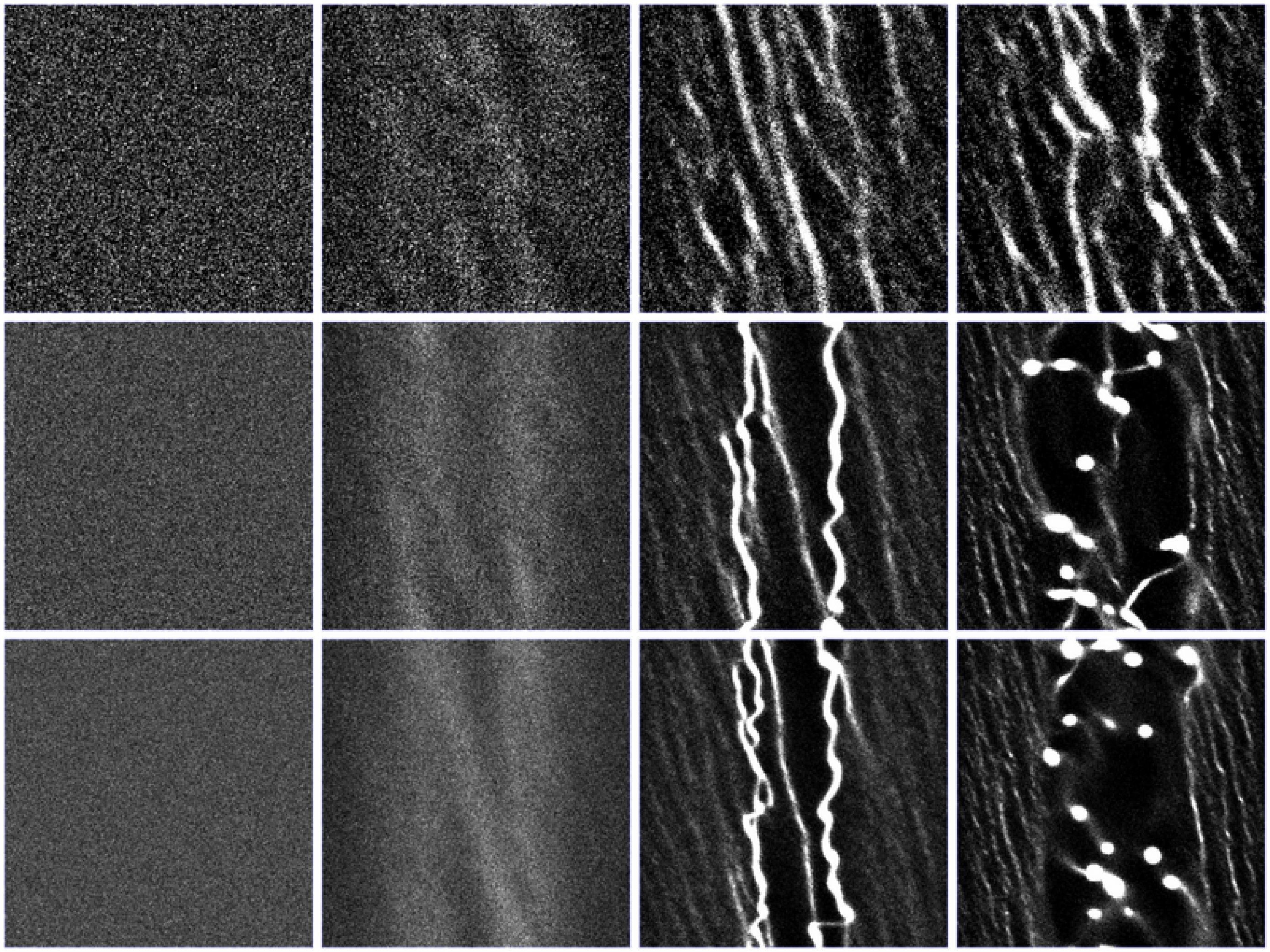}
\caption{Scaled particles: Snapshots of the particle distribution in the $xy$ plane. The simulations (from top to bottom) use 40~000, 320~000, 640~000 particles with their physical size given by Eq.~\ref{eq:anum}. In all simulations, the collision timescale was kept constant. The snapshots were taken (from left to right) at $t=0, 30, 37, 40\,\Omega^{-1}$. The simulation with 40~000 particles has a large filling factor by the last frame, which prevents clumps from forming. The intermediate resolution simulation (middle row) and the highest resolution simulation (bottom row) have more and smaller particles and thus a smaller filling factor. The results (i.e., number of clumps in the last frame) are very similar in the intermediate and high resolution simulations.   \label{fig:snapcollisions}}
\end{figure*}

Snapshots of the particle distribution of two simulations are shown in \Fig \ref{fig:snapsmoothing}. 
Both simulations use 160~000 particles and all parameters, except the smoothing length, are the same. 
The top row is the simulation with a smoothing length given by Eq.~\ref{eq:anum}, whereas the bottom row uses a smoothing length that is ten times larger. 
The simulations correspond to the blue (medium dashed) and red (solid) curve in \Fig \ref{fig:veldisp_smooth}. 
We show the snapshots to illustrate the importance of the smoothing length in simulations without physical collisions. 
One can see that the simulations differ already before clumps form. 
Stripy structures appear on a scale given by Eq.~\ref{eq:toomrelength}. As soon as we enter the unstable regime at~$t\sim32\Omega^{-1}$ (i.e., the velocity dispersion begins to rise, see also \Fig \ref{fig:veldisp_smooth}) the simulations evolve very differently. The simulation on the top row of \Fig \ref{fig:snapsmoothing} forms many clumps at an early time, whereas the bottom row simulation forms only a few, more massive clumps at later times. 

This can be confirmed by looking at spectra of the same two simulations as shown in \Fig \ref{fig:spectr_smooth}. 
These spectra were generated by mapping the particles onto a $128\times 128$ grid in the $x-y$ plane and computing the fast Fourier transform of the mapping in the $x$ direction. 
The resulting spectra were finally averaged in the $y$ direction \citep[see][for a complete description of the procedure]{Tanga2004}. 
As suggested by the snapshots, the nonlinear dynamics of the gravitational instability strongly depends on the smoothing length used.
In particular, one observes that smaller scales are amplified more slowly for larger smoothing lengths. 
The resulting spectra at $t=40\Omega^{-1}$ also differ significantly. 
With a small enough smoothing length, the spectrum looks almost flat, whereas a large smoothing length introduces a cutoff at $k/2\pi\simeq 10$. 
We also note that the smoothing length in the latter case is of the order of the grid size ($k/2\pi\sim100$). 
The cutoff observed clearly demonstrates that the smoothing length modifies the dynamics on scales up to 10 times larger than the smoothing length itself.

\subsection{Scaled particles}

The velocity dispersion evolution of simulations with $N=160\,000, 320\,000, 640\,000$ particles including physical collisions is presented in \Fig \ref{fig:veldisp_scaled}. In these runs, no gravitational smoothing length is needed. Again, the simulations start from a stable equilibrium and after $t=30~\Omega^{-1}$, we change the same parameters as in the non-collisional case to push the system into the gravitationally unstable regime.
Snapshots of the particle distributions for three runs are also plotted in \Fig \ref{fig:snapcollisions}. The middle and bottom rows of snapshots correspond to the purple (medium dashed) and dark blue (long dashed) lines of \Fig \ref{fig:veldisp_scaled}.

One can see in \Fig \ref{fig:veldisp_scaled} as well as \Fig \ref{fig:snapcollisions} that the intermediate and high resolution runs produce very similar results. We call those simulations converged, as a change in particle number does not change the outcome. 
An additional, lower resolution run ($N=40\,000$) is not converged because the filling factor is too large in dense regions, preventing clumps from being gravitationally bound. 
The problem does not occur in the non-collisional cases because particles are allowed to overlap. 
Since the filling factor scales with particle number as~$1/\sqrt{N_\text{num}}$, this issue is resolved for runs with more than a few hundred thousand particles. 
We note that the velocity dispersion might differ slightly at later stages for the converged runs because the final clump radius is still determined by the physical particle radius and therefore by the particle number. 

\begin{figure}[htpH]
\centering
\includegraphics[angle=270,width=\columnwidth]{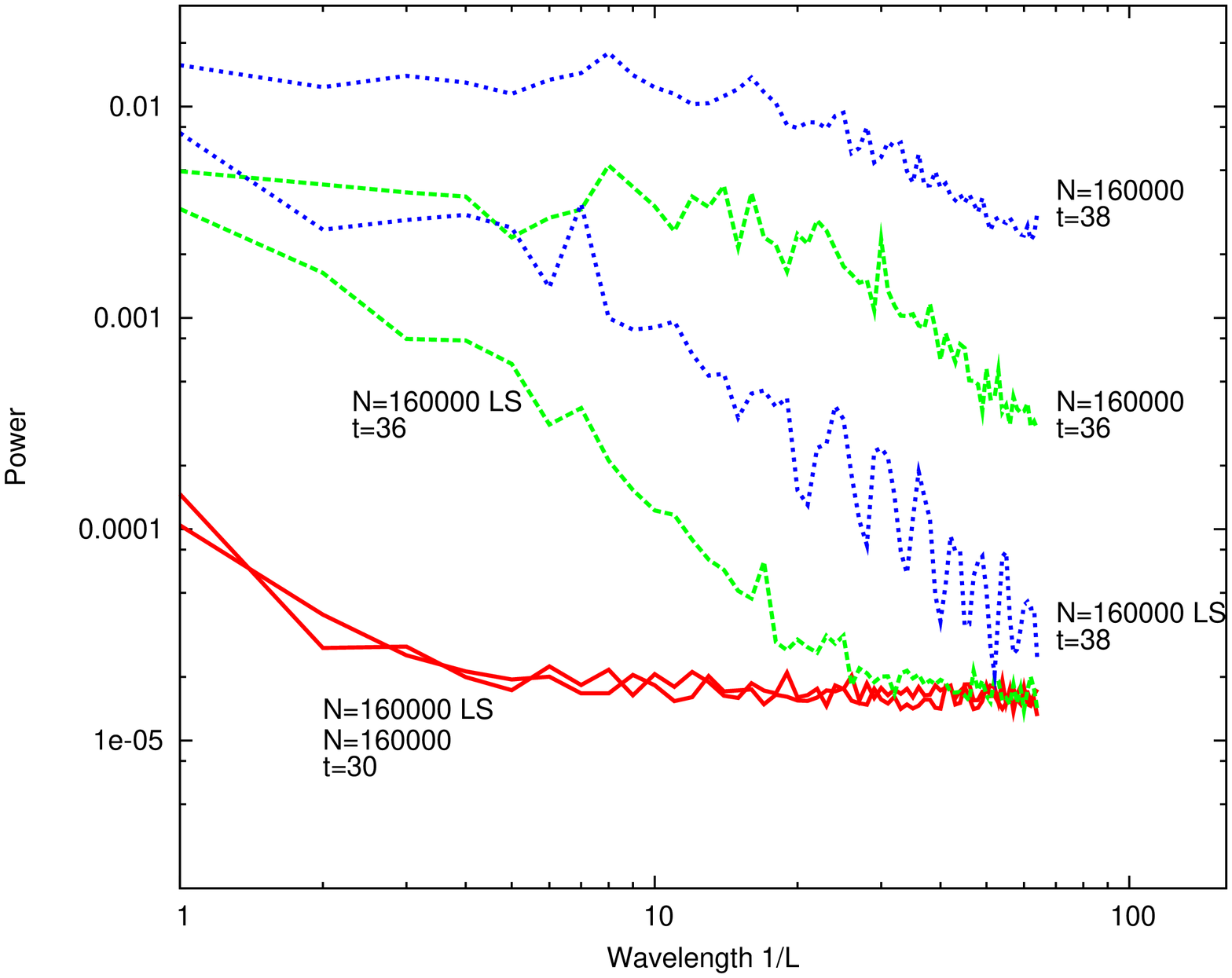}
\caption{Super-particles: Density distribution spectrum with 160~000 particles and two different smoothing lengths that differ by a factor 10 at times $t=30$ (red, solid curve), $t=36$ (green, long dashed curve), and $t=38$ (blue, short dashed curved). The spectra at similar times are not on top of each other because the simulations are not converged.\label{fig:spectr_smooth}}
\centering
\includegraphics[angle=270,width=\columnwidth]{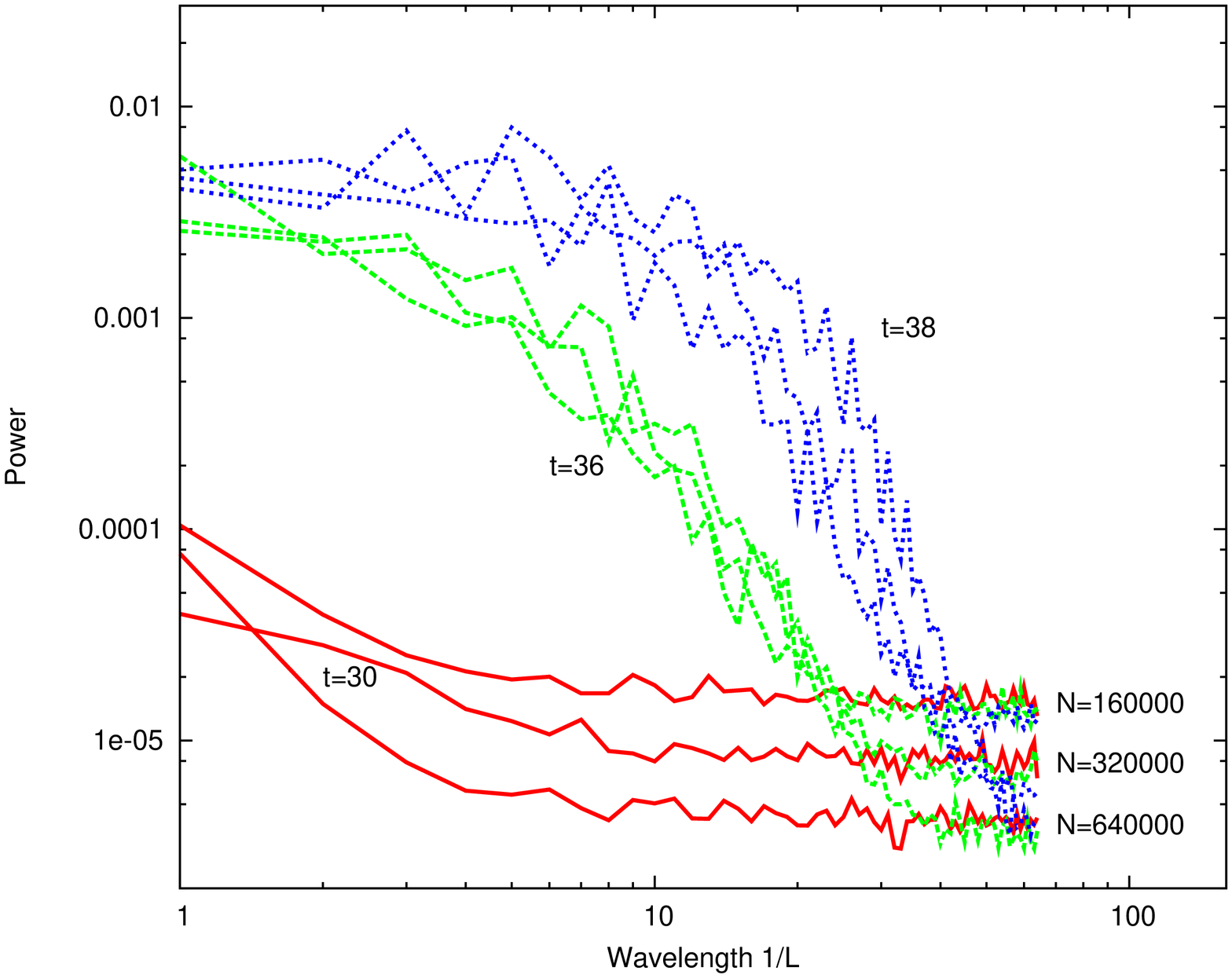}
\caption{Scaled particles: Density distribution spectrum with 160~000, 320~000, and 640~000 particles (colors are equivalent to \Fig \ref{fig:spectr_smooth}). At each time, the spectra are converged, i.e., the spectra are on top of each other and independent of the particle number, besides the noise level on very small scales. At late times ($t\gtrsim38$, blue curve), one begins to see more structure on small scales in simulations with larger number of particles (see text).  \label{fig:spectr_coll}}
\end{figure}

We show the density distribution spectrum in \Fig \ref{fig:spectr_coll} as a function of the number of particles. The resulting spectra do not depend on the number of particles in the system, as expected. In comparison with \Fig \ref{fig:spectr_smooth}, the simulations with collisions are converged since the dynamic is not controlled by the numerical parameter $N$, which is the only free parameter in the system. 

At very late times ($t\gtrsim 38\Omega^{-1}$), the spectra begin to show a systematic trend towards more structure on smaller scales for runs with larger $N$. This is expected and due to the filling factor, already mentioned above. As soon as one clump in the simulation is only determined by the size of the individual particles it consists out of, our approach breaks down.
One can also see these clumps forming by looking at the velocity dispersion in \Fig \ref{fig:veldisp_scaled}. The particles inside a clump begin to dominate the velocity dispersion over the background after $t\sim38\Omega^{-1}$. A clear indication of this is the spiky structure with a typical correlation time of $\sim\Omega^{-1}$ corresponding to different clumps interacting and merging with each other.  One way around this issue, which will be considered in future work, is to allow particles to merge \citep{Michikoshi2009}. Using this \textit{accretion model}, the mass of the clump can be used as an upper limit.

\section{Discussion and implications}\label{sec:disc}
In this paper, we have shown that convergence in $N$-body simulations of planetesimal formation via gravitational instability can be achieved when taking into account all relevant physical processes. 
It is absolutely vital to simulate gravity, damping, excitation, and physical collisions simultaneously, as the related timescales are of the same order and therefore all effects are strongly coupled. 

A set of simulations is defined to be converged when the results do not depend on the particle number or any other artificial numerical parameter such as a smoothing length.
As a test case, a box with shear periodic boundary conditions was used, containing hundreds of thousands of self-gravitating \mbox{(super-)particles} in a stratified equilibrium state. 
The particles were then pushed into a self gravitating regime that eventually led to gravitational collapse.
Simulations with and without physical collisions were studied for a range of particle numbers to test convergence.
 
In cases without physical collisions, convergence can not be achieved. However, it was possible to change multiple free simulation parameters at the same time, namely the particle number $N$ and the smoothing length $b$, such that in special cases the results did not depend on the particle number. We note however that there is a free parameter in the simulation (i.e., the smoothing length) that effects the outcome and makes it impossible to find the \textit{real} physical solution. 

In the protoplanetary disc, physical collisions dominate over gravitational scattering. In the case of planetesimal formation, the system is marginally collisional. Physical collisions will partially randomise the particle distribution, whereas a smoothing length does not because the softening length is very large compared to the gravitational cross-section. Even if the gravitational particle-particle scattering becomes important, it can be seen from Eq.~\ref{eq:crossgravity} that the velocity dependence of the cross-section $\sigma_G$ differs fundamentally from the velocity independent cross-section $\sigma_C$. We therefore argue that the super-particle approach should not be used when collisions become important.

With collisions, the simulation outcome is both quantitatively and qualitatively very different from simulations with no physical collisions. The initial size of the clumps is larger and the number of clumps is smaller. As there is no smoothing length, there is effectively one fewer free parameter. Changing the particle number while keeping the collision time constant does not change the outcome. We therefore call these simulations converged. 

Additional tests including inelastic collisions with a normal coefficient of restitution of $0.25$ have been performed to confirm that the results do not depend on the way physical collisions are modelled. As expected, the qualitative outcome in terms of number of clumps and clump size is different because the physical properties of the system have been changed. However, once individual collisions are resolved the results converge in exactly the same manner as in those simulations presented above (which have a coefficient of restitution of $1$). 

We have also explored other gravity solvers by comparing a fast Fourier (FFT) gravity solver with the BH tree code (using a smoothing length) and confirmed our previous results. 
The grid in the FFT code acts as a smoothing length and a direct comparison between the two gravity solvers gave an approximate effective smoothing length of a quarter of the grid length.
Because one has to solve individual particle-particle interactions, the grid size has to be smaller than the physical size of the particles. As a result, the FFT method is always slower than a tree code. 
In addition, the tree structure can be reused to search efficiently for collisions.

This paper focused on the numerical requirements to study gravitational instability and the formation of planetesimals in protoplanetary discs. 
The initial conditions were chosen such that the equilibrium is a well defined starting point for the convergence study. The collision, damping, collapse and orbital timescales are all of the same order, which is effectively the worst case scenario. 
To provide any constraints on planetesimal formation itself, one would have to properly simulate the turbulent background gas dynamics which is beyond the scope of this paper.

However, we were able to determine the numerical resolution needed to resolve dust particles in a protoplanetary disc.
Assuming that one wishes to simulate dust particles realistically, including collisions, and that the particles are uniformly distributed in a box of base $L^2$ and height $H$, the size of each particle is determined by the collision rate in the real disc (see Eq.~\ref{eq:anum})
\begin{eqnarray}
a_{\text{num}} = \sqrt{\frac{N}{N_{\text{num}}}} a,
\end{eqnarray}
where $a$ and $N$ are the size and number of particles in the volume $L^2H$ of the real disc. 
The number of particles in the simulation $N_{\text{num}}$ has to be large enough to ensure that the filling factor is low until clumps form. 
The requirement that the filling factor is less than unity at $t=0$ is given by
\begin{eqnarray}
\frac43 \pi  \left(a_{\text{num}}\right)^{3} \cdot N_{\text{num}} \le C_f\;  L^2 H,
\end{eqnarray}
where $C_f$ is a safety factor. Although $C_f\lesssim1$ would be enough to resolve collisions initially, it is insufficient to simulate the collapse phase. During the collapse, the filling factor rises rapidly. Assuming that one wishes to resolve collisions correctly when the particles have contracted by one order of magnitude, one has to include a safety factor of $C_f=10^{-3}$. If the collapse occurs mainly in two dimensions, as in the simulations presented here, a factor of $C_f=10^{-2}$ is sufficient. 
Furthermore, one has to ensure that the box contains at least a few unstable modes (see Eq.~\ref{eq:toomrelength}). All this together places tight numerical constraints on numerical simulations of planetesimal formation via gravitational instability.
 
In future work, we will expand the discussion to include more physics, namely a proper treatment and feedback of the background gas turbulence. This will allow us to further clarify the behaviour of planetesimals in the early stages of planet formation and eventually test the different formation scenarios.

\begin{acknowledgements}
We thank J. Papaloizou, G. Ogilvie, S.-I. Inutsuka, A. Johansen, Y. Lithwick, and A. Youdin for helpful discussions and comments. Hanno Rein is supported by an Isaac Newton Studentship and St John's College Cambridge. Geoffroy Lesur acknowledges support from STFC. Zoë M. Leinhardt is an STFC postdoctoral fellow. Simulations were performed on the Astrophysical Fluids computational facilities and on Darwin, the Cambridge University HPC facility.
The authors are grateful to the Isaac Newton Institute for Mathematical Sciences in 
Cambridge where the final stages of this work were carried out during the Dynamics of Discs and Planets research programme. A simplified version of the simulation used in this paper can be downloaded free of charge as an application for the iPhone at \url{http://itunes.com/apps/gravitytree}.
\end{acknowledgements}
\bibliography{full}
\bibliographystyle{aa}
\end{document}